**Enhancing student learning of two-level quantum systems with interactive simulations**

Antje Kohnle[a)], Charles Baily, Anna Campbell, Natalia Korolkova

*School of Physics and Astronomy, University of St Andrews, St Andrews, KY16 9SS, United Kingdom*

Mark J Paetkau

*Department of Physical Sciences, Thompson Rivers University, Kamloops, Canada V2C, OC8*

**ABSTRACT**

The QuVis Quantum Mechanics Visualization project aims to address challenges of quantum mechanics instruction through the development of interactive simulations for the learning and teaching of quantum mechanics. In this article, we describe evaluation of simulations focusing on two-level systems developed as part of the Institute of Physics Quantum Physics resources. Simulations are research-based and have been iteratively refined using student feedback in individual observation sessions and in-class trials. We give evidence that these simulations are helping students learn quantum mechanics concepts at both the introductory and advanced undergraduate level, and that students perceive simulations to be beneficial to their learning.

**I. INTRODUCTION**

Quantum mechanics holds a fascination for many students, but learning quantum mechanics is difficult. The counterintuitive behaviour of quantum systems often disagrees with our classical and common-sense ideas, leading to student difficulties that arise when classical thinking is applied to quantum systems.[1-16] Quantum phenomena typically cannot be observed directly and are far-removed from everyday experience. Complicated mathematics is required to describe even simple phenomena. Instruction often focuses on particularly simple abstract and idealized systems that are mathematically tractable, but may not help learners make real-world connections to quantum phenomena and acquire corresponding physical intuition.

One approach recently gaining favour is to introduce quantum theory using so-called two-level or two-state systems.[17-21] Examples of such systems are a single photon that can be found in two distinct beams in an interferometer, a spin ½ particle which can be found in a "spin up" or "spin down" state along a given axis, and a two-level atom with a ground state and only one excited state. In each case, linear superpositions of the two states are also possible. These systems are isomorphic in that they are described by the same mathematical formalism. Such two-level systems are physical realisations of a quantum bit or qubit, having two distinguishable states and superpositions between them.

Developing the theory using two-level systems can have multiple advantages: It immediately immerses students in the concepts of quantum mechanics by focusing on experiments with quantum superposition states that have no classical explanation. It allows from the start a discussion of interpretations of quantum mechanics and recent applications such as quantum information technology. It can be mathematically less challenging than the continuum wave mechanics approach, requiring only basic linear algebra instead of calculus and differential equations.

In this article, we describe the evaluation of research-based interactive simulations with accompanying activities to support quantum mechanics instruction using two-level systems. These simulations are part of the QuVis Quantum Mechanics Visualization project[22], and can be freely accessed for use online or download from the QuVis website www.st-andrews.ac.uk/physics/quvis (the "New Quantum Curriculum sims" collection). The simulations (17 in total) in this collection cover the topics of linear algebra, fundamental quantum mechanics concepts, single photon interference, the Bloch sphere representation, entanglement, hidden variables and quantum information. Simulations are suitable for a first course in quantum physics, although a subset of simulations (six in total) require complex numbers and two simulations require the manipulation of 2×2 matrices.

These simulations are embedded in a full curriculum as part of the Institute of Physics (IOP) Quantum Physics resources at quantumphysics.iop.org. The IOP resources include around 80 short articles centred on questions written by researchers in quantum information and foundations of quantum mechanics, as well as the simulations and accompanying activities, problems and a glossary of terms.[20] The IOP website is free to use but requires users to register. This allows the site to save users' difficulty ratings of articles, which are shown via color codes in a navigation panel.

This article is organized as follows: in section II, we describe features of interactive simulations that address challenges of quantum mechanics instruction and make them useful tools for learning. Section III describes evaluation outcomes from studies of student learning at the introductory and advanced undergraduate level. Simulations were used in courses as collaborative computer classroom activities and as homework assignments, to learn new concepts as well as consolidate concepts. Section IV has conclusions and outline future plans.

## II. INTERACTIVE SIMULATIONS AS INSTRUCTIONAL TOOLS

Interactive simulations are powerful tools for the learning of quantum mechanics. They can make the invisible visible, and thus give students insight into microscopic processes that cannot be directly observed. They can help students make connections between different representations, such as physical, mathematical and graphical representations of phenomena.[23-25] They can reduce complexity to focus on key ideas by depicting idealized and simplified situations compared with actual laboratory experiments, and thus reduce cognitive load by eliminating extraneous material.[26] They allow students to compare and contrast different situations, such as comparing the behaviour of classical waves and single photons under the same experimental conditions.

They allow students to carry out experiments to see how quantum-mechanical quantities are determined experimentally. They can also challenge students' classical ideas by allowing them to assess whether classical models can correctly predict experimental outcomes.[20]

The QuVis simulations on two-level systems build on principles of effective multimedia and interaction design such as breaking complex content into parts, using multiple representations, and implicit scaffolding through interaction design.[27] They complement other research-based collections of quantum mechanics simulations[22,25,28-31] by providing content on topics not otherwise available or not available at this level.

In what follows, we describe two simulations that will be referred to in section III. Figure 1 shows a screenshot of the *Phase shifter in a Mach-Zehnder interferometer* simulation, which allows students to send single photons through an interferometer and to insert a phase shifter to vary the relative phase between the two arms.

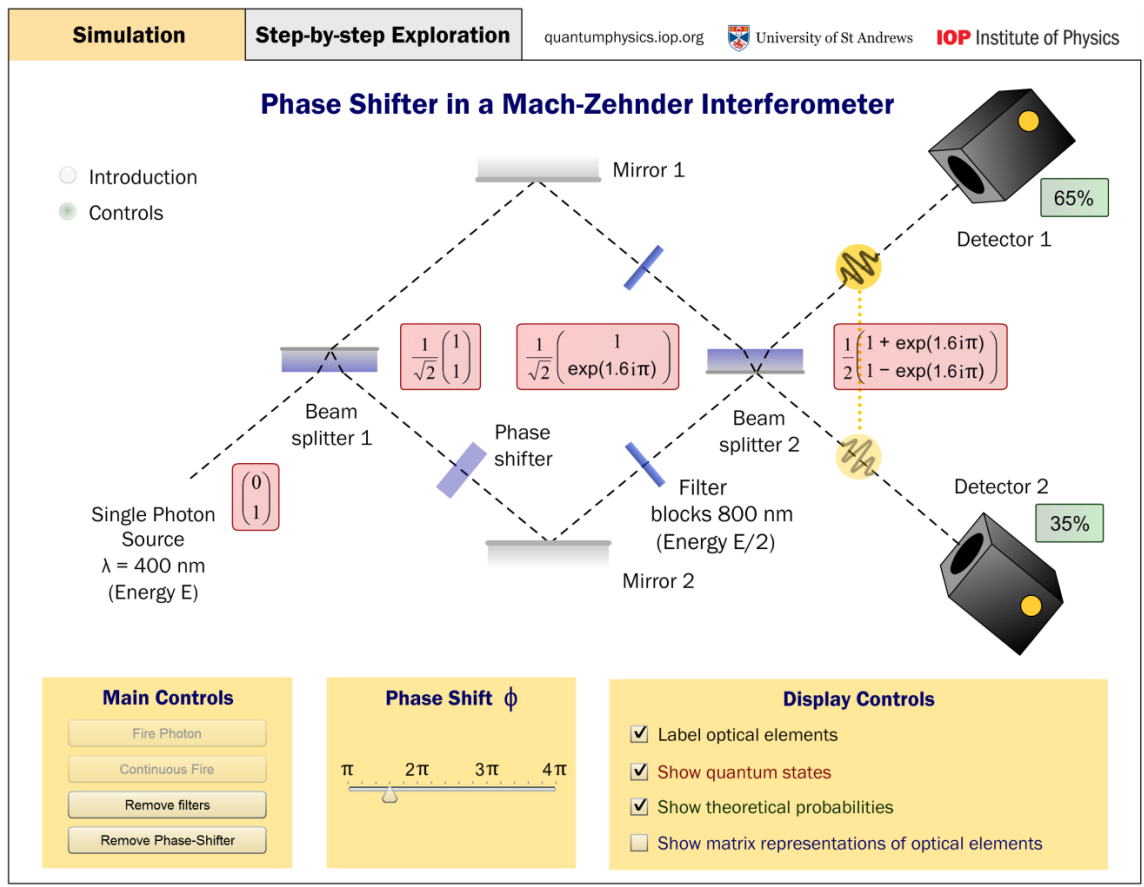

FIG. 1. A screenshot of the *Phase Shifter in a Mach-Zehnder Interferometer* simulation. This simulation aims to help students make connections between physical and mathematical representations of single photon interference and the measurement process.

The simulation depicts single photons and the photon superposition state in order to help students develop productive mental models of single photon interference. This photon visualization is the outcome of a study investigating the impact of different visualizations on student understanding of quantum superposition.[32] Students can insert filters that block half the energy of the source photon to test ideas about photon superposition, e.g. the incorrect idea of quantum superposition being akin to a classical object splitting into two half-energy components. Students can display the quantum state at various points in the interferometer and the mathematical representations of the optical components, to help make connections between the physical setup and mathematical representations. Detection events are depicted as flashes in the detectors.

Figure 2 shows a screenshot of the *Superposition states and mixed states* simulation. This simulation allows students to use a Stern-Gerlach apparatus that can be oriented along two orthogonal axes to investigate whether they can experimentally distinguish mixed states and superposition states.

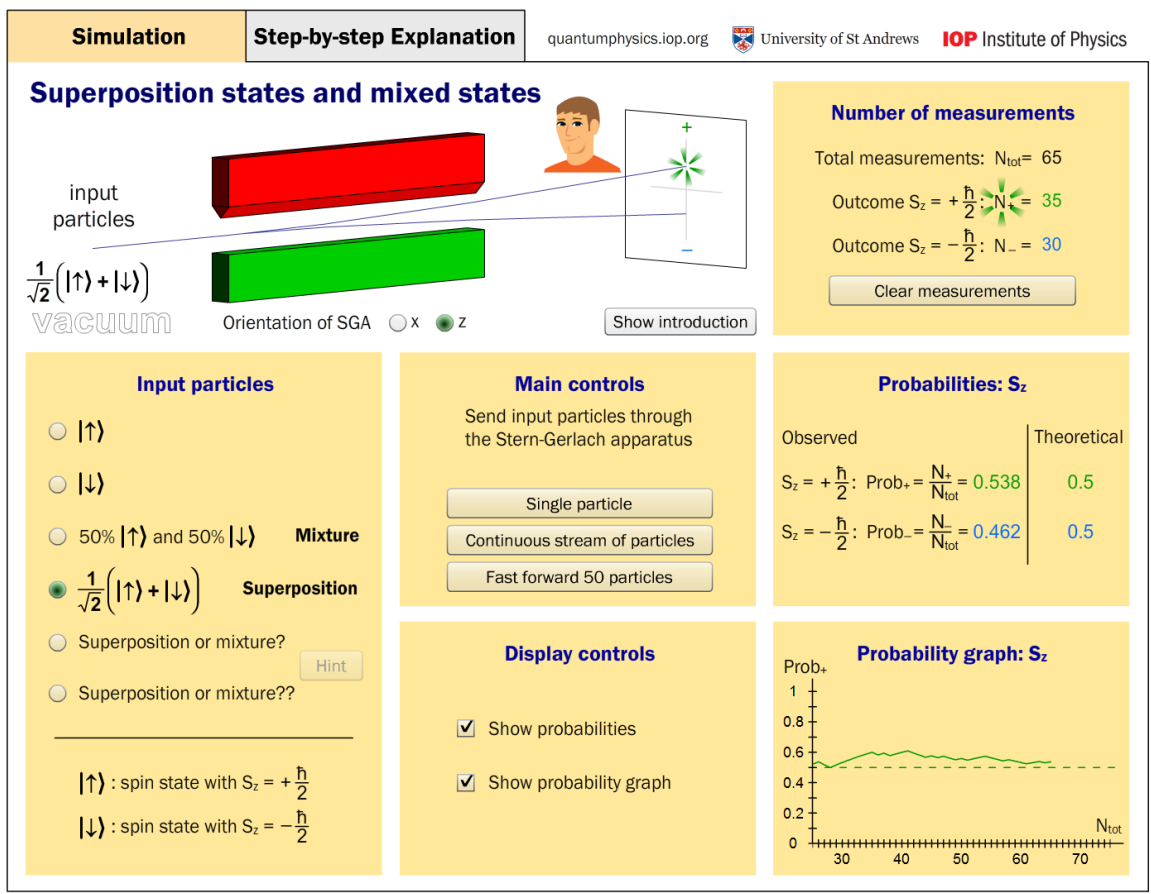

FIG. 2. A screenshot of the *Superposition states and mixed states* simulation. This simulation allows students to investigate whether they can experimentally distinguish mixed states from superposition states.

Students can choose different input spin states, which include a superposition state, a corresponding mixed state and two unknown input states (see the "Input particles" panel in Fig. 2). The simulation shows the individual input spin states and spin measurement outcomes, and the experimentally determined and theoretical outcome probabilities mathematically and graphically.

In the *Superposition states and mixed states* simulation, flashes are used to help students make connections between the measurement outcome on the screen and the outcomes shown in the "Number of measurements" panel. Color is used to help students differentiate between the two measurement outcomes. Color is also used to link the number of measurement outcomes to their respective observed and theoretical detection probabilities as shown mathematically and graphically (the three right-hand panels in Fig. 2).

## III. EVIDENCE FOR STUDENT LEARNING

For the 17 simulations in the "New Quantum curriculum sims" collection, in total we carried out 42 hours of observation sessions with 19 student volunteers, 17 of whom were from the introductory level. In these individual observation sessions, students first interacted freely with a simulation while thinking aloud and describing what they were making sense of and what they found confusing. Students then worked on the activity associated with the simulation. Afterwards, students completed a survey on their experiences of using the simulation and suggestions for improvement. These sessions lasted for two hours, and students typically explored two simulations and in some cases three simulations during this time. Each simulation was used by typically 2 to 4 students. For a number of simulations and activities, minor changes based on our observations were implemented prior to testing them again with subsequent students. All simulations were refined based on outcomes from these sessions.

We have so far carried out in-class trials at the University of St Andrews with 9 of the simulations. Five simulations have been used in an introductory Quantum Physics course taken by students in their first or second year at university. Four simulations have been used in an Advanced Quantum Mechanics course for senior level students. Four of the simulations at the introductory level were used as 50-minute long collaborative computer classroom activities. The other in-class trials used simulations and activities as homework assignments. Table I in section C gives details of seven of the simulations used and the number of students completing the simulation assignments. We also used the *Phase shifter in a Mach-Zehnder interferometer* simulation at the introductory level and the *Graphical representation of complex eigenvectors* simulation at the advanced level, but only a small fraction of the class completed the assignment and these are therefore not included in Table I. In this section, we give examples showing evidence of student learning from the observation sessions and the in-class trials.

## A. Promoting student-driven inquiry

Well-designed interactive simulations promote engaged exploration, where students actively explore and make sense of the phenomena shown led by their own questioning. Careful design of simulations in terms of affordances (actions that are available) and constraints (features that restrict actions) can make simulations effective through implicit scaffolding of students' exploration, guiding students without them feeling guided.[26] They can also promote scientific abilities such as setting up experiments, data-handling, the ability to work with multiple representations, the scientific method of generating, testing and falsifying hypotheses, using analogy to reason about phenomena, and model-building based on experimental outcomes.

From observing students explore the simulations freely, we find evidence that the simulations engage these students in making connections between different quantities, making predictions, testing them and interpreting the outcomes. In what follows, we give an example of student-driven inquiry from an individual student observation session using the *Phase shifter in a Mach-Zehnder interferometer* simulation (see Fig. 1). In the transcript reproduced below, the student is changing the "Phase Shift" slider (shown in the bottom middle of Fig. 1) and exploring how this effects the detection probabilities in detectors 1 and 2. The student did not have the tick boxes "Show quantum states" or "Show theoretical probabilities" checked during this sequence, both of which are displayed in Fig. 1 for clarity. Also, the filters shown in Fig. 1 were not inserted excepting at the start. The student is freely exploring the simulation without an associated activity.

> *Let's shoot something through first [clicks on "Fire Photon" button] – that's interesting. (laughs)*
>
> *[clicks "Insert filters"] Currently the filter does not appear to be having any effect. Let's remove the filters and insert a phase shifter instead. [removes the filters, inserts the phase shifter, per default the phase is at π]*
>
> *The detector at which they arrive at has switched.*
>
> *Ah, you can alter the phase. [changes the slider for the phase shifter to 1.6π]*
>
> *That's interesting. Ah yes, so, umh, again these two are connected. [Points with mouse to the dashed photon superposition just before reaching the detector]. One is slightly brighter than the other suggesting that the probability of them arriving at detector 1 is greater than at detector 2. That does seem to be the case as they pass through – there seems to be a little bit more in detector 1 than in detector 2.*
>
> *[moves phase shift to 2π] I guess this will go back to detector 1 as you would suspect. And again with 4π. [moves phase shift to 4π]*

*[moves to 3π] If there's an integer number of ... an odd number of π produces a wave going directly to detector 2, an even number produces a photon heading directly to detector 1 and then in between [moves slider to 2.5π] sort of the probability slowly gradually shifts from detector 1 to detector 2.*

In this short sequence, the student is making sense of the visualizations as shown in Fig. 1, interpreting the dotted line between the two components of the superposition to suggest they are connected, and the transparency of the photon to depict the detection probability. The student is making predictions and testing them experimentally, e.g. that changing the phase shift by $2\pi$ leaves the detection probabilities unchanged. The student is generalizing results to come up with general rules about the influence of the phase shift on the detection probabilities.

**B. Learning gains**

In what follows, we describe evidence for student learning using the *Superposition states and mixed states* simulation shown in Fig. 2. Initial trials of this simulation in individual student observation sessions showed that students had some difficulty understanding the mixed state. We revised the simulation so that the state of each input particle is shown in the graphics window. Thus, for the mixed state, the revised simulation displays a random sequence of spin-up and spin-down input states. This allows students to make connections between the input state and the measurement outcome seen as a flash on the screen. For the superposition state, the revised simulation displays a sequence of identical input states. We found that this revision helped students make sense of the difference between mixed states and superposition states. We also swapped the order of mixed states and superposition states in the input panel, as we found that students found it easier to make sense of the measurement outcomes along both axes for the mixed state compared with the superposition state.

Studies show that including small puzzles or challenges in simulations can encourage prolonged engagement and inquiry.[23,26] To be productive, challenges need to be aligned with the learning goals of the simulation. Challenges should require both high behavioural activity (interaction with the simulation) and high psychological activity (cognitive processing of content). The *Superposition states and mixed states* simulation (Fig. 2) includes two unknown input states labeled with question marks. These challenges are aligned with the learning goals of the simulation. Thus, we can assess whether students have achieved the learning outcomes by measuring success in solving these built-in challenges. A hint button in the simulation tells users to determine the measurement outcome probabilities along the two axes. The activity to the *Superposition states and mixed states* simulation asks students whether they encounter similar mixtures of objects in their everyday experience, and helps students make sense of the difference between quantum superposition and classical mixtures. The activity gets students to explore the known input states before asking them to solve the challenges.

After revisions from the observation sessions were incorporated, we used the *Superposition states and mixed states* simulation in an Advanced Quantum Mechanics course for senior level students. This course includes Hilbert space, the matrix formalism, pure and mixed states via the density matrix, entanglement and quantum information processing. The simulation activity was given as a homework assignment and collected for marking one week later. Students were given a short pre-test in the class on the day the assignment was given, and a short post-test in the class on the day they handed in their simulation activity responses. Students only received feedback on their answers after the post-test. The homework assignment, pre- and post-tests did not count towards the course grade. There were 33 students in the class, of whom 20 completed pre-test, homework and post-test. The pre-test question was as follows, with choice b) being deemed to be the correct answer:

> *In what follows, $|\uparrow\rangle$ and $|\downarrow\rangle$ refer to spin states where the z-component of spin $S_z = +\hbar/2$ and $S_z = -\hbar/2$ respectively.*
>
> *Consider*
>
> *A) a random mixture of spin ½ particles, where each particle is either in state $|\uparrow\rangle$ or in state $|\downarrow\rangle$ with on average 50% of each type.*
>
> *B) spin ½ particles each in the same superposition state $1/\sqrt{2}(|\uparrow\rangle + |\downarrow\rangle)$.*
>
> *Imagine you had a large number of spin ½ particles of the random mixture described in A), and a large number of particles each in the superposition state described in B). Which of the following statements is/are true concerning these particles?*
>
> *a) The cases A and B may look different, but there is no way they can be distinguished experimentally.*
>
> *b) If we measure a different component of spin than $S_z$, we can experimentally distinguish between the cases A and B.*
>
> *c) The difference between the cases A and B is just in our knowledge of the system. Particles of type B are actually in the state $|\uparrow\rangle$ or $|\downarrow\rangle$; we just do not know which of these states each particle is in.*
>
> *d) Particles of type B actually oscillate in time rapidly between being in state $|\uparrow\rangle$ and being in state $|\downarrow\rangle$. This is why we measure the particle to be either $|\uparrow\rangle$ or $|\downarrow\rangle$ when we measure the z-component of spin.*

The post-test question was identical except a changed order of the choices. Students were asked to rate their confidence in their answer as "Certain", "Somewhat certain", "Somewhat uncertain" or "Uncertain" and explain their reasoning. The course itself only discussed mixed states in terms of the density matrix and not at the conceptual level of the simulation.

Figure 3 shows the fraction of students that succeeded in solving the built-in challenges. 16 of 20 students (80%) correctly identified the mixture and 15 of 20 students (75%) the correct fractions in the mixture. 15 of 20 students (75%) correctly identified the superposition state and 14 of 20 students (70%) the correct coefficients in the superposition state. Given that these are advanced level students, one could surmise that this success in completing the built-in challenges could be due to prior knowledge and not due to learning through interacting with the simulation. The pre- and post-test responses allow us to test this hypothesis.

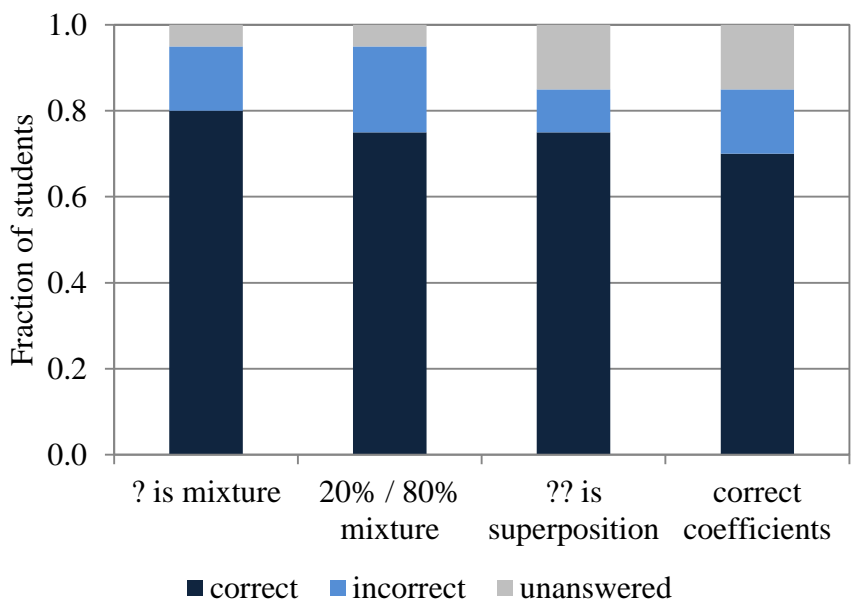

FIG. 3. The fraction of students that succeeded in determining the two unknown input states in the *Superposition states and mixed states* simulation. 20 students in total completed the homework assignment using this simulation.

We coded pre- and post-test responses as correct, partially correct and incorrect. Initial codes were based on students' choices, with students choosing more than one answer which included the correct answer b) being coded as partially correct. These codes were then refined using the student reasoning, which all students had completed. If a student chose both a) and b), but explained that a) was correct if only the z-component of spin is considered, this answer was coded as fully correct. Incorporating the reasoning led to two responses being coded differently.

Figure 4 shows the outcomes for the pre-test (left) and the post-test (right) in terms of the percentage of students with correct, partially correct and incorrect answers. The most common incorrect answer on the pre-test was choice a). Choices c) and d) were only chosen by 1 and 2 students respectively. The colors in Fig. 4 denote the certainty ratings. One can see that the fraction of students choosing the correct response substantially increased between the pre- and the post-test, and that students were on average more certain of their response on the post-test.

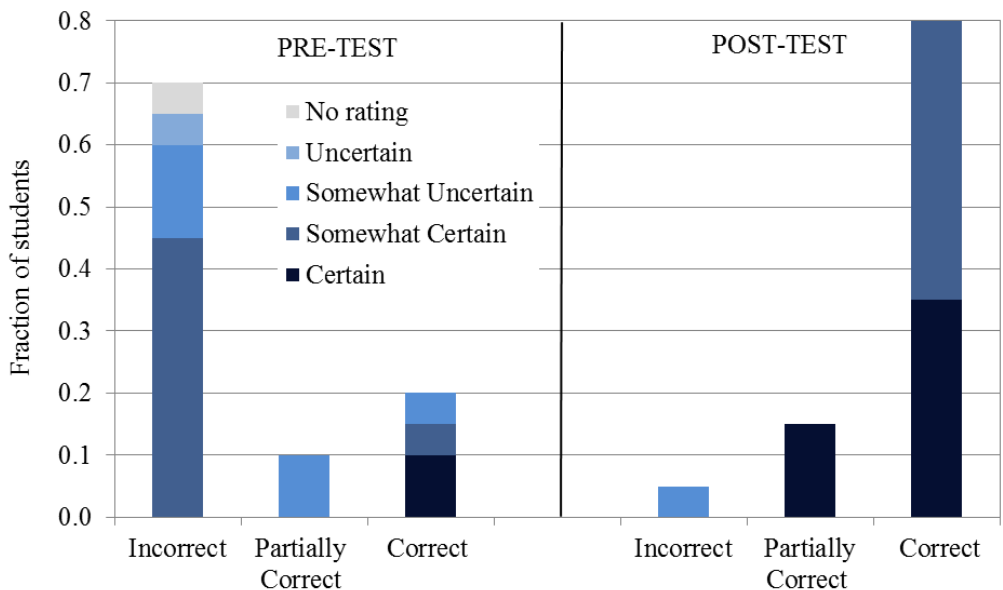

FIG. 4. Outcomes for the pre-test (left) and post-test (right) responses for the *Superposition states and mixed states* simulation. 20 students in total completed pre-test, post-test and the simulation activity.

## C. Student perceptions

Interactive simulations can empower students to learn through a safe and enriched learning environment where equipment is not breakable. They can empower students through the interactive elements, which allow cycles of trial-and-error exploration through immediate feedback on actions. A comparative study we carried out in our 2013/14 introductory Quantum Physics course illustrates the impact of interactivity on student engagement. One representative group of students (those with a particular lab day, N=34) worked on an activity using printed screenshots from the simulation *Entangled spin ½ particle pairs versus an elementary hidden variable theory* in a 50-minute classroom session. The other group of students (N=48) worked on the same activity using the actual simulation in a computer classroom, again for a 50-minute session. At the end of these sessions, all students completed a survey on their experiences and suggestions for improvement. Figure 5 shows the responses for the two groups to a survey question on how enjoyable students found the activity on a Likert scale from 1 (not enjoyable) to 5 (very enjoyable). One can see a marked difference in the distributions, with students overall finding it more enjoyable to work with the simulation. For the group that used the printed screenshots, 19 of 22 comments in total to the survey question "Do you have any suggestions for improvement" related to the usefulness of interactivity. Examples of typical student comments are "Much easier to play around with simulations so that you can run tests and experiments." and "I think it's more straightforward to use the computer for this as you are able to adjust the sliders and see directly how this changes the result of the experiment." All students had used QuVis simulations on two-level systems in the course prior to this study.

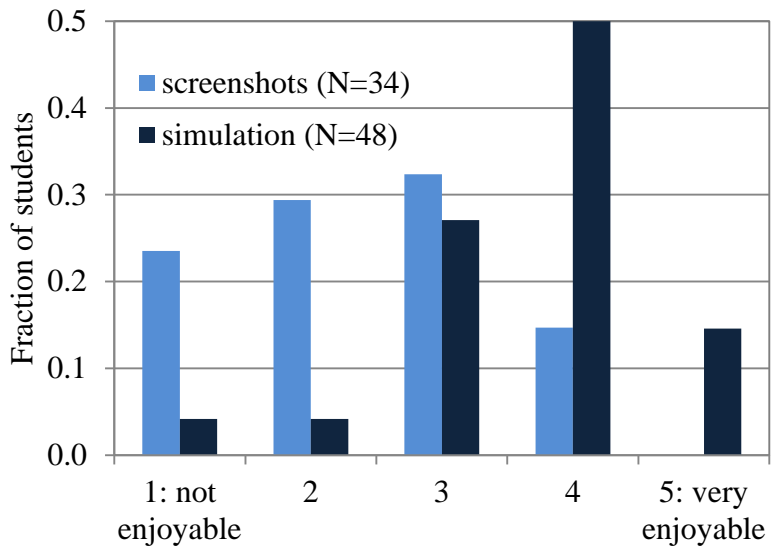

FIG. 5. Outcomes from a comparative study of students working on the same activity with printed screenshots of a simulation (N=34) and the simulation itself (N=48). The histogram shows student responses to the question how enjoyable they found the activity on a Likert scale from 1 (not enjoyable) to 5 (very enjoyable).

Students' perceived usefulness of the simulations in improving their understanding can impact their engagement with these resources. Table I shows outcomes from surveys that students completed directly after working with a simulation for in-class trials in 2013 and 2014 at the introductory and the advanced level. For two of the in-class trials, only a small fraction of students completed the survey, so these results are not reproduced here. On the whole these results are positive, with the majority of students across both levels finding simulations useful in improving understanding.

For the 13/14 session, we asked students (N=73) in the introductory Quantum Physics course on an end-of-course survey *"How useful for learning quantum physics have you found the simulations used in the course?"*. 40% of students stated "very useful" and 43% "useful", and no students stated the simulations were not useful. In total, 5 simulations were used in this course.

TABLE I. Student perceptions of the usefulness of simulations in improving their understanding on a Likert scale from 1 (not useful) to 5 (very useful). Shown are the averages and standard deviations (in parentheses) to this question for seven in-class trials of simulations from the two-level collection.

| Simulation | Level used | Number of students | Usefulness in improving understanding |
| --- | --- | --- | --- |
| Interferometer experiments with photons, particles and waves | introductory | 62 | 4.3 (0.6) |
| The Expectation Value | introductory | 72 | 4.2 (0.7) |
| Entangled spin ½ particle pairs versus an elementary hidden variable theory | introductory | 48 | 4.0 (0.7) |
| Entangled spin ½ particle pairs versus local hidden variables | introductory | 52 | 4.0 (0.9) |
| Superposition states and mixed states | advanced | 21 | 3.8 (0.8) |
| Entanglement: The nature of quantum correlations | advanced | 14 | 4.5 (0.9) |
| Quantum key distribution | advanced | 14 | 4.4 (0.5) |

## IV. CONCLUSIONS AND FUTURE PLANS

We have used the QuVis simulations on two-level systems in both introductory and advanced undergraduate quantum physics courses as collaborative computer classroom activities and as homework assignments. Students at both levels perceive the simulations to benefit their learning. We have initial evidence from observation sessions and in-class trials that simulations are helping students learn quantum mechanics concepts, including topics such as entanglement and hidden variables at the introductory level that are typically covered only at the advanced level. However, further multi-institutional evaluation studies are needed to ensure simulations are useful to students from a wide range of backgrounds.

Some of the in-class trials have pointed to particular issues. For example, *The Expectation Value* simulation allowed introductory level students to successfully learn this concept just from the simulation, but post-test responses to a question differentiating the expectation value and the most likely value for a single measurement were only moderately successful. For the *Quantum key distribution with entangled spin ½ particles* simulation used with introductory level students to learn about quantum cryptography just from the simulation, student responses to the activity and post-test questions were well answered, with the exception that it was not always clear how the two observers check for errors. We are currently incorporating changes into these simulations specifically targeting difficulties found.

The simulations described here were originally coded using Adobe Flash. The recoding of simulations using HTML5 is currently ongoing, and a number of simulations in the collection are already available in the new format on the QuVis website. The HTML5 simulations run on both desktop computers as well as tablet-based devices. We are incorporating revisions based on our in-class trials during the recoding process. Future work includes the development of further simulations on two-level systems, in particular with entangled photon pairs and with a focus on quantum information. We also plan to develop more open and exploratory activities, including intrinsically collaborative activities that require students to bring together their individual contributions. We plan to optimize these activities using observation sessions where students work collaboratively using the simulations.

ACKNOWLEDGEMENTS

We thank the outstanding undergraduate physics students who coded these simulations. We thank all students taking part in the observation sessions and the in-class trials. We thank the UK Institute of Physics for funding the two-level simulation development, and for developing and maintaining the quantumphysics.iop.org website. We thank Gina Passante from the University of Washington for helping with the development of pre- and post-tests for the simulations.ACKNOWLEDGEMENTS

We thank the outstanding undergraduate physics students who coded these simulations. We thank all students taking part in the observation sessions and the in-class trials. We thank the UK Institute of Physics for funding the two-level simulation development, and for developing and maintaining the quantumphysics.iop.org website. We thank Gina Passante from the University of Washington for helping with the development of pre- and post-tests for the simulations.